\documentclass[review]{elsarticle}
\usepackage{lineno}
\usepackage{geometry}
\usepackage{booktabs}
\usepackage{threeparttable}
\usepackage{amsmath}
\usepackage{amsthm}
\usepackage{mathrsfs}
\usepackage{amssymb}
\usepackage{graphics}
\usepackage{graphicx}
\usepackage{multirow}
\usepackage{subfig}
\usepackage{bm}
\usepackage{algorithm}
\usepackage{algpseudocode}
\usepackage[colorlinks,linkcolor=red, anchorcolor=blue,citecolor=green]{hyperref}

\modulolinenumbers[1]
\graphicspath{{./figs/}}

\DeclareMathOperator*{\argmin}{arg\,min}

\begin{document}
\begin{frontmatter}
\title{A Spatial Deconstruction Behavioural Model for Pedestrian Dynamics} %

\author{Yao Xiao} 
\address[0]{School of Intelligent System Engineering, Sun Yat-Sen University, Shenzhen, China}
\cortext[mycorrespondingauthor]{E-mail address: xiaoyao9@mail.sysu.edu.cn} 

\begin{abstract}
How to reproduce realistic motion in simulations has always been a fundamental problem for pedestrian dynamics, and a critical challenge for current studies is the natural correlation of the movement choices and the human behaviours. To advance the study, we deconstruct the two-dimensional space of pedestrian crowds with a behavioural motion network on basis of Voronoi diagram and Delaunay diagram, and accordingly formulate a behavioural model considering both the long-term route choice and the real-time velocity choice. The comparisons between simulation results and real-world movements suggest that the behavioural model can not only deal with traditional scenarios but also be capable of describing well-agreed movements in challenging conflicting situations. The spatial deconstruction idea provides a geometric perspective to recognize and reproduce the pedestrian behaviours, and it can further benefit the understanding and realization of pedestrian dynamics and even robot navigation. 
\end{abstract}

\end{frontmatter}

\section{Introduction} \label{section introduction}

Considering the growing frequency of public activities and the increasing congestion extent at daily peak commuting times, the investigation of pedestrian crowds has attracted the attention of many researchers. These studies have benefited the design of public facilities \citep{Helbing2005} as well as the organization of crowd activities \citep{Bain2019}. Among these researches, a fundamental topic is about how to reproduce realistic pedestrian motion behaviours, especially in complex situations.

Different types of microscopic modeling ideas\citep{Helbing1995,Blue2001,Moussaid2011,VanDenBerg2011,Alexandre2016,Xiao2016} have been proposed to simulate pedestrian crowds in a two-dimensional space. The widely-used social force model \citep{Helbing1995,Helbing2000} regards pedestrians as Newton's Second Law based particles driven with a combination of a driving force (the desire to reach a target), social forces (the untouched psychological interaction) and contact forces (the touched physical interaction). Through reviewing the traditional model, a heuristic model \citep{Moussaid2011} further introduces two simple cognitive procedures based on the distance of obstructions to adapt the pedestrian directions and speeds. Meanwhile, other modeling idears such as the cellular automota model\citep{Blue2001,Fu2018}, velocity obstacle model\citep{VanDenBerg2011,Douthwaite2019}, machine learning method\citep{Alexandre2016,Ma2019}, Voronoi model\citep{Xiao2016} are proposed and attempted. In spite of the reproduction of well-agreed fundamental diagrams and many self-organized phenomena, human behaviours are still not properly considered in current models. First, a real pedestrian considers not only the perceptible real-time velocity choice but also the latent long-term route choice among dynamic pedestrians. The lack of route choice consideration would lead to an imperfect motion mechanism and create an insuperable gap between simulation and reality (e.g., the lack of roundabout strategies in complex situations). Second, current models rely too much on particle properties (e.g, individual repulsion) rather than directly introducing realistic pedestrian motion strategies (e.g., following other pedestrians or making a detour). For these problems, how to deconstruct the flexible pedestrian space and find the corresponding behaviour-based route choice is the critical issue.

Herein, a motion network based on the Voronoi diagram and its dual graph Delaunay diagram \citep{Fortune1987} is proposed. Through a combination between the geometric features of the motion network and the behavioural characteristics of pedestrian dynamics, a series of systemic and behavioural routes are recognized and defined. Accordingly, a behavioural model considering both the long-term route choice and the real-time velocity choice is proposed.

The rest of paper is organized as follows. Section \ref{section model} introduces the behavioural motion network and describes the formulation of the behavioural model. In Section \ref{section results}, simulations of both traditional and challenging scenarios are conducted. Section \ref{section discussion} discusses the model results and the potential improvements in the future.

\section{Model} \label{section model}

\begin{figure}[!ht]
\centering{\includegraphics[width=1\textwidth]{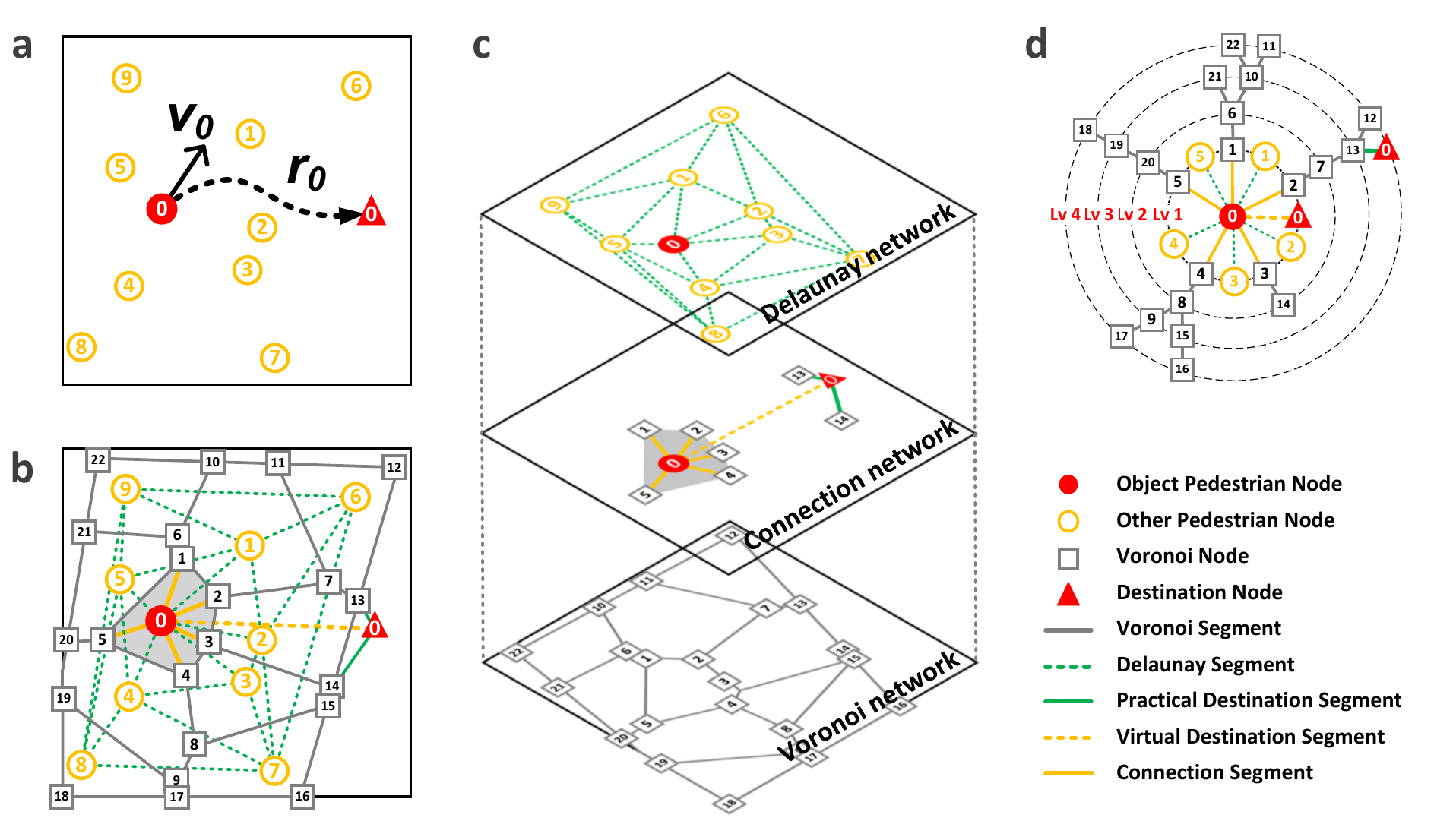}}
\caption{\textbf{Pedestrian dynamics problem and motion network.} \textbf{a}, Sketch map of pedestrian motion problem. 10 pedestrians represented by circles are initialized in a closed square space. The object pedestrian $P_0$ needs to leave for the corresponding destination $D_0$, and his/her supposed velocity choice $v_0$ and route choice $r_0$ are shown by solid and dashed arrows, respectively. \textbf{b}, Pedestrian motion network. \textbf{c}, Sub-networks of pedestrian motion network. \textbf{d}, Feasible routes of object pedestrian. Note that these feasible routes are obtained by finding the minimum time cost and considering the route segment counts.}\label{figroutechoice}
\end{figure}

The modeling of pedestrian dynamics is usually simplified into a two-dimensional space, and the core problem (see Eq. \ref{eqMotionStrategy}) for pedestrian $P_i$ is about how to reach the destination $\bm{D}_i$ while considering geometries $G$ and pedestrians ($P_1, P_2 , \ldots, P_n$).

\begin{equation} \label{eqMotionStrategy}
\bm{s}_i = (\bm{r}_i, \bm{v}_i) = f(P_1, P_2 , \ldots, P_n; \bm{G}; \bm{D}_i)
\end{equation}
Inspired by the practical pedestrian decision process, the motion strategy $\bm{s}_i$ of pedestrian $P_i$ is divided into two phases, the long-term route choice $\bm{r}_i$ which predicts the anticipatory motion strategic route in the foreseeable future, and the real-time velocity choice $\bm{v}_i$ which indicates the present velocity direction and speed (Fig.\ref{figroutechoice}-a). Namely, the route choice guides the instantaneous velocity choice, while the velocity choice reflects the anticipatory route choice. In a former work \citep{Xiao2016}, the decision process of the real-time velocity choices is initially formulated. Here, the original real-time model is extended, and the long-term route choice and the real-time velcoity choice are integrated together.

\subsection{Long-term route choice}

The spatial deconstruction can contribute to the recognition of the pedestrian status and the topological relationship among pedestrians, and the long-term route choice behaviours can also be significantly benefited. For this reason, a pedestrian motion network (see Fig.\ref{figroutechoice}-b) is formulated at each time step based on the Voronoi diagram of the pedestrians and its dual graph Delaunay diagram.

\textbf{Motion network}. The motion network contains four types of nodes and five types of segments, and it can be divided into three sub-networks (see Fig.\ref{figroutechoice}-c). The Delaunay network describes a kind of topological relationship \citep{Ballerini2008} between one pedestrian and his/her adjacent pedestrians (e.g., the neighbors of pedestrian $P_0$ include pedestrians $P_1$, $P_2$, $P_3$, $P_4$, and $P_5$), and the potential pedestrian following direction pattern can be accordingly inspired. The Voronoi network indicates the intermediate passageway between two pedestrians (e.g., Voronoi segment $N_2N_7$ represents the intermediate passageway of pedestrian $P_1$ and $P_2$), and the potential pedestrian crossing, overtaking and detour pattern can be obtained. The connection network links the pedestrian, the destination and the Voronoi nodes, as well as the Voronoi network and the Delaunay network. On basis of the integrated motion network, a series of routes can be found from a pedestrian to his/her destination.

\textbf{Perception scope}. Generally, it is impractical for a pedestrian to schedule a tortuous route in a large crowd, and an overly long planned route might not make sense in the time-varying dynamic surroundings. As a result, the real perception scope of a pedestrian is usually limited, and it can be implemented by introducing a threshold value of route length. Three types of route choices among the potential routes are found and defined according to the first segment of route, i.e, destination pattern (virtual destination segment), following pattern (Delaunay segment) and detour pattern (connection segment). Since only the detour pattern is considered to be an aggressive behaviour in the model, all the segment lengths of the routes except for the detour routes are limited to be only one.

In practical, a pedestrian generally precisely calculates the cost in the perception scope and roughly estimates the potential cost to the goal outside the perception scope. Therefore, the A* search algorithm \citep{Hart1968} is introduced and modified as Eq. \ref{eq route choice 1} to find the optimal strategic route $\bm{r}_i^*$ of pedestrian $P_i$ from the feasible route set $\bm{R}_i$ as shown in Fig. \ref{figroutechoice}-d.

\begin{equation}\label{eq route choice 1}
\bm{r}_i^* = \underset{\bm{r}_{i}\in \bm{R}_{i}}{\mathrm{\argmin}}\left( \frac {\delta^r} {v_i^*} \cdot (\sum_{k=1}^K \delta^s L_i^k + \sigma_i D_{\bm{r}_i})  \right)
\end{equation} 
where $L_i^k$ and $D_{\bm{r}_i}$ respectively represent the length of $k$th route segment inside the perception scope and the Euler distance from the tail node of the strategic route to the destination outside the perception scope. $v_i^*$ indicates the speed determined according to the first segment of route, and $\sigma_i$ is a detour coefficient (see \ref{section appendix detour coefficient}). $\delta^s = {\rm exp}(\sum_{m=1}^{M^s} \phi_m \lambda_m^s)$ is a penalty term for the segment, and $\phi_m$ is the type judgment term (i.e., if the segment meets the required type, $\phi_m$ = 1, otherwise, $\phi_m$ = 0). $\delta^r = {\rm exp}(\sum_{m=1}^{M^r} \lambda_{m}^r)$ is a penalty term for the strategic route.

In our model, two types of segment penalties and one type of route penalty are taken into consideration. For the Delaunay segment which indicates the following behaviour, the synergy effect with the following target is formulated as, $\lambda_1^s = \alpha_1 \cdot ((1-\bm{e}_i\cdot\bm{e}_i^f) + (1-\bm{e}_i^0\cdot\bm{e}_i^f))$, where $\bm{e}_i^0$, $\bm{e}_i$ and $\bm{e}_i^f$ respectively indicate the unit vector of the pedestrian desired direction, the pedestrian velocity direction and the following target's velocity direction. The Voronoi segment indicates the passageway between two adjacent pedestrians from a geometrical view. To reflect the difficulty level of the across behaviour, the time penalty is formulated as, $\lambda_2^s = \alpha_2 \cdot  {d_i^r} / {w^2}$, where $d_i^r$ and $w$ are respectively the radius of pedestrian circle and the width between the adjacent pedestrians. The type of route penalty is proposed to realize the conflict avoidance behaviour, $\lambda_1^r = \beta_1 \cdot n(A), A = \{ \bm{x}_{ob}| \ \|\bm{x}_{ob}- \bm{x}_c \|<2d_i^r, \bm{x}_{ob} \in \bm{X}_{ob} \}$, $\bm{x}_{c}$ is the location of tail node of current route, and $\bm{X}_{ob}$ contains the existing tail nodes of other strategic routes. The other parameters here are respectively set as,  $\alpha_1 = 1$, $\alpha_2 = 3$ and $\beta_1 = 5$. Note that further terms can also be added in the route and segment penalty formula to describe the related impacts. 


\subsection{Real-time velocity choice}

\begin{figure}[!ht]
\centering{\includegraphics[width=0.7\textwidth]{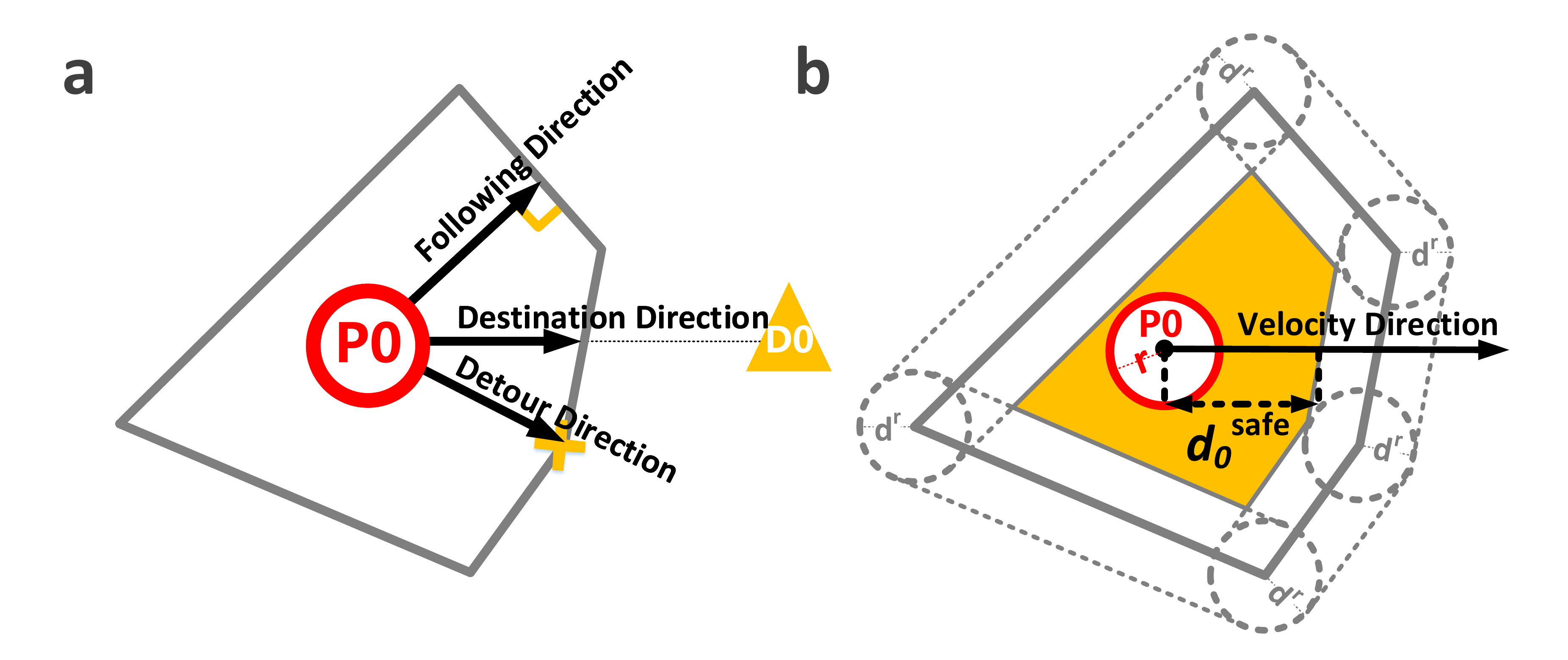}}
\caption{\textbf{Sketch map of Voronoi cell and velocity choice.} \textbf{a}, Direction choices. In the Voronoi cell of pedestrian $P_0$, the three types of direction choices including, destination direction, following direction and detour direction are marked. \textbf{b}, Safe distance and speed determination.} \label{figvelocitychoice}
\end{figure}

The process of the real-time velocity choice for pedestrian $P_i$ can be divided into two parts \citep{Moussaid2011}, the determination of direction $\bm{e}_i^*$ and the determination of speed $v_i^*$. 

\textbf{Direction choice}. The three types of the first segments in route choice just correspond to three direction choice behaviours, respectively. In a Voronoi cell (see Fig. \ref{figvelocitychoice}-a), the Delaunay segment perpendicular to the corresponding Voronoi cell boundary denotes the direction to adjacent pedestrians which is a general following direction $\bm{e}_{follow}$. The connection segment pointing to the Voronoi node represents the direction between two adjacent pedestrians which is theoretically an across or detour direction choice $\bm{e}_{detour}$. The virtual destination segment directly corresponds to the direction choice to the destination $\bm{e}_{dest}$. In summary, the three alternative direction choices for the optimal direction are introduced as $ \bm{e}_i^* \in \{ \bm{e}_{dest}, \bm{e}_{follow}, \bm{e}_{detour} \}$, and our experimental investigations \citep{Xiao2018} show that the combination of the three direction choices is able to describe most motion behaviours in pedestrian crowds. The pedestrian velocity direction is determined according to the long-term route choice, and the direction unit vector is $\bm{e}_i^* = \bm{e}(\bm{r}_i^*)$.

\textbf{Speed choice}. Since the corresponding Voronoi cell of each pedestrian (e.g., the gray Voronoi cell area in Fig.\ref{figroutechoice}-b) contains all of the points closer to that pedestrian than to any other, it is defined as the personal space. Based on this, the safe distance $d^{safe}$ is defined as the maximum distance to the boundary of the personal space along the velocity direction (see Fig. \ref{figvelocitychoice}-b where the pedestrian is represented by a circle). Accordingly, the speed of pedestrian $P_i$ is determined as, $v_i^* = {\rm min}(v_i^0, 2(d_i^{r}+d_i^{safe})/\tau_i)$, where $v_i^0$ is the desired speed. In addition, a velocity obstacle concept \citep{Fiorini1998, VanDenBerg2011} can be applied to further improve the practical conflict avoidance performance.

A second-order (acceleration) motion equation (Eq. \ref{eq velocity}) is introduced \citep{Helbing1995,Helbing2000} to describe the velocity change of pedestrian $P_i$ at time step $t$.

\begin{equation}\label{eq velocity}
\frac {{\rm d}\bm{v}_i(t)}{{\rm d}t} = \frac{\bm{e}_i^* \cdot v_i^* - \bm{v}_i(t)}{\tau_i},
\end{equation}
In the equation, the relaxation time is set as $\tau_i = {\rm ln}(1+2d_i^r+2d_i^{safe})$, hence the pedestrians can be more responsive in critical situations.

\section{Results} \label{section results}


The spatial deconstruction based behavioural model can reproduce realistic crowd movements in different situations. In the section, we firstly test the model performance with several traditional indexes and then explore the model practicality in the challenging circle antipode experiments.  

\subsection{Traditional scenarios}

\begin{figure}[!ht]
\centering{\includegraphics[width=0.7\textwidth]{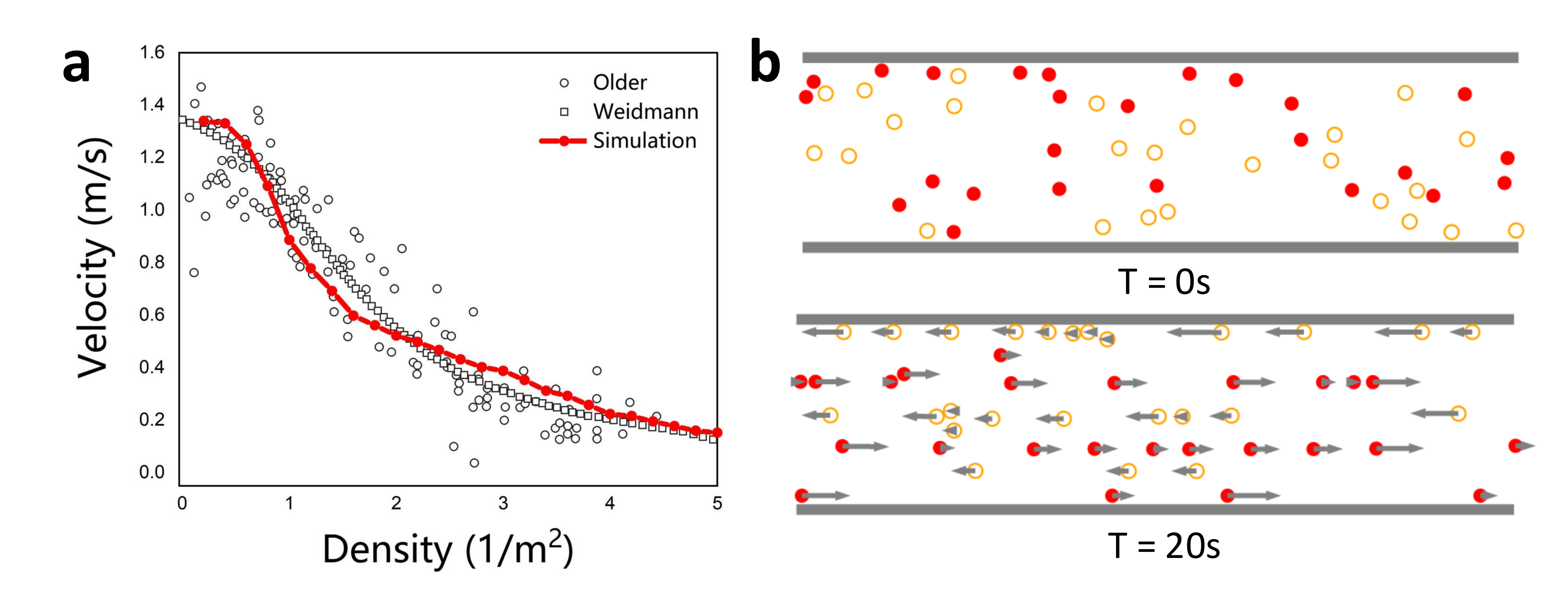}}
\caption{\textbf{Simulation performance in traditional scenarios.} Pedestrians are randomly initialized in a $15 \times 5$ m corridor with periodic boundaries, and the desired velocity is normally distributed as $v^0 \sim N(1.35, 0.5)$. \textbf{a}, Velocity-density relation in the uni-directional flow. The simulation results are obtained by averaging the pedestrian velocities, and each data point responds to a certain participant number. \textbf{b}, Lane formation in the bi-directional flow. Half of the participants (orange hollow circles) head toward the left end of the corridor and the rest (red solid circles) head for the other side. The arrows represent the velocity directions.} \label{figtradresults}
\end{figure}

The fundamental diagram \citep{Seyfried2005} mainly describes the relationship between macroscopic parameters (e.g. density, velocity and flow), and it is currently a most widely-used indicator \citep{Boltes2019} to validate the pedestrian model performance. Fig. \ref{figtradresults}-a shows the well-agreed velocity-density curve pattern between the simulated results and the experimental results \citep{Older1968, Weidmann1993}.

Another critical model reliability indicator is related to the reproduction of the self-organized phenomena \citep{Moussaid2011,Helbing2005,Helbing2009}, among which the lane formation \citep{Seyfried2005,Helbing2005} in bi-directional flow is a representative one. The model reproduces obvious crowd lanes in the bi-directional flow scenarios after a simulation of 20 s as shown in Fig.\ref{figtradresults} - b. Besides, more variety of the self-organized phenomena (e.g., stop-and-go wave, and arching phenomenon) as well as critical situations (e.g., U-shape obstacle, and aggressive pedestrian) can be captured by the behavioural model.


\subsection{Circle antipode Scenarios}

\begin{figure}[!ht]
\centering{\includegraphics[width=1\textwidth]{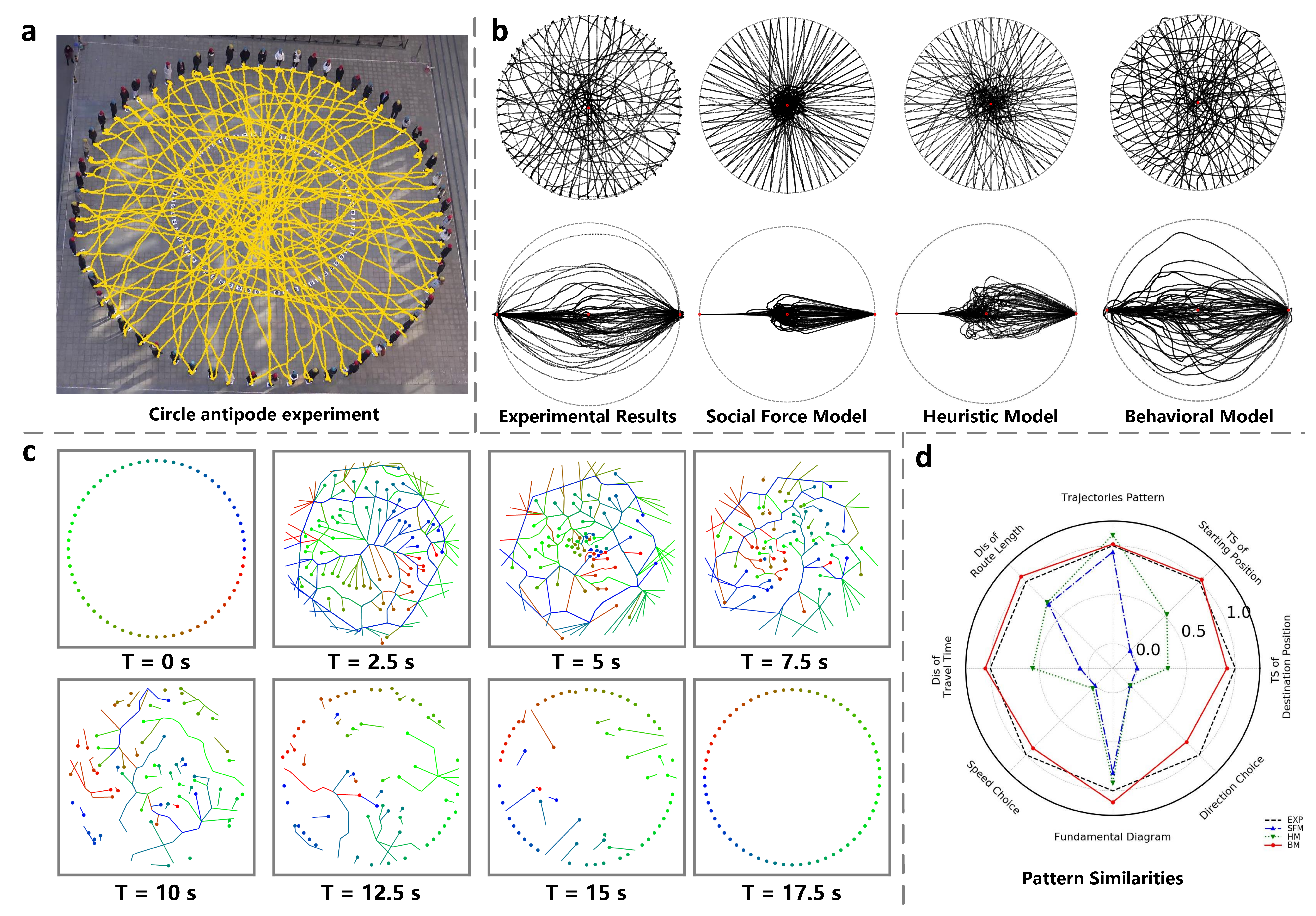}}
\caption{\textbf{Circle antipode experiments and simulations.} \textbf{a}, Empirical trajectories of a sample circle antipode experiment. The yellow lines represent the ground trajectories during the experiments. \textbf{b}, Trajectory comparisons among experiments and different models. Considering the symmetric experimental condition, the original trajectories above are further rotated to the same starting position on the left side and the same destination position on the right side. \textbf{c}, Snapshots of the behavioural model simulation. \textbf{d}, Pattern similarities between experiments and simulations. A total of 8 indexes are introduced for evaluation and more details can be found in \cite{Xiao2022}. The data necessary to support the findings of this manuscript are available in a public repository (\url{https://github.com/rickyspy/Behaviour-Model-Data})} \label{figcirexpsims}
\end{figure}

The route navigation and conflict avoidance behaviours in crowds are challenging topics in the simulations of pedestrian dynamics. To have an ideal research scenario, the circle antipode experiments in which the pedestrians are uniformly initialized on the circle and required to leave for the antipode position simultaneously are designed. On 22 May 2019, we organized a series of repeated circle antipode experiments at the square before a teaching building in Beijing Jiaotong University. A total of 64 participants are included in the experiments, and they are all students aged between 18 to 28 and height between 170 cm to 180 cm. On top of the teaching building, a high-resolution camera is set to record the whole experiment, and PeTrack \citep{Boltes2013,Boltes2010} is then applied to recognize the pedestrians in the video and obtain the corresponding trajectories on the ground (see Fig.\ref{figcirexpsims}-a). In the circle antipode experiment, all the 64 participants are initialized on a circle of 10m radius. After a short pause when all participants respectively reached their destination, the participants are simultaneously required to leave for the antipode position again, and the process repeated 20 times except for several warm-up experiments. There are two significant characteristics in the experiments that would benefit the model validation. First, the shortest routes of pedestrians converge on the center position of the circle, so the central region is likely to be a conflicting area. This knowledge can significantly benefit the research of route navigation and conflict avoidance behaviours. Second, except for the pedestrian heterogeneities, the motion conditions for each participant are symmetric in the experiments (e.g., symmetric starting position, symmetric destination position, and symmetric surrounding). Thus, it's a kind of repeat experiment for the participants (i.e., 64 repeated samples are obtained in the experiment with 64 participants), and the indeterminate features of the pedestrian behaviours can be greatly overcome. 

In Fig.\ref{figcirexpsims}-b, the original trajectories as well as the rotated trajectories (rotated to the same starting point) are extracted and processed. It's easy to find that the pedestrian route choices especially the detour choices have been made at the very beginning when there is no actual conflicts in the way to the destination. Namely, the intelligent individuals in our experiments can not only deal with the conflicts at present but also predict the future potential conflicting situation among pedestrian crowds. Considering the fundamental mechanism of traditional models, it's too challenging to reproduce well-agreed pedestrian behaviours in the circle antipode experiments (see the trajectories of the social force model \citep{Helbing1995,Helbing2000} and heuristic model \citep{Moussaid2011} in Fig.\ref{figcirexpsims}-b.

On the contrary, the proposed behavioural model can be used for the extremely conflicting situations and reproduce well-agreed trajectory patterns as shown in Fig.\ref{figcirexpsims}-b. From further snapshots of the behavioural model simulation in Fig.\ref{figcirexpsims}-c (including pedestrian and his/her corresponding route choice), we can find two major types of route choices. Some just walk through the conflicting central region and the others tend to detour from the relaxed region on the outside. In fact, the heterogeneous types of route choices in the simulation are consistent with the empirical results.

To further quantitatively measure the pattern similarities between simulations and experiments, we introduce a comprehensive evaluation framework containing 8 sets (4 types) of indexes (see Fig. \ref{figcirexpsims}-d). The pattern similarities of repeated experiments(EXP), social force model(SFM)\citep{Helbing2000,Helbing1995}, heuristic model(HM)\citep{Moussaid2011} and our behavioural model(BM) are investigated within the evaluation framework. Note that there is also a pattern similarity among repeated experiments due to the indeterminacy of pedestrian behaviours, and each index is normalized with the empirical results in repeated experiments. As a result, a greater pattern similarity indicates a better performance, and a score of $1$ theoretically indicates a similar performance as the repeated empirical experiments. In Fig. \ref{figcirexpsims}-d, a slight advantage of the heuristic model is found for the trajectories pattern index, and the side choice behaviour at the beginning of the circle antipode experiments is believed to be the critical cause \citep{Xiao2022}. Nevertheless, our behavioural model shows a great advantage to reproduce realistic results in the circle antipode experiments for most indexes.

\section{Discussion} \label{section discussion}


The proposed pedestrian behavioural model can not only deal with the existed conflicts and make practical avoidance, but also anticipate the potential conflicting areas in the future and schedule the route in advance. The combination of the long-term route and real-time velocity choices on basis of the motion network formulates the foundation of these complicated pedestrian walking behaviours. In addition to agreeing closely with the traditional simulation results, the behavioural model can reproduce more reliable crowd movements in the challenging circle antipode experiment which requires macroscopic level motion planning.


This work provides a geometric spatial deconstruction approach to understand and combine pedestrian behaviours, and both the route and velocity choice modules on basis of the spatial deconstruction approach can be further developed. For instance, it's not hard to find more modeling possibilities based on the current velocity determination module which contains three discrete direction choices (i.e., destination direction, detour direction, and following direction). Also, the current route choice module in our model is still simplified, and more route planning approaches based on the formulated motion network is likely to be practicable and promising. 

The proposed spatial deconstruction behavioural model can contribute to the understanding and development of robot navigation \citep{Khambhaita2020,Pandey2017}. What's more, the widely-applied deep learning methods show great potential for predicting pedestrian trajectories\citep{Alexandre2016,Zhang2019}. Incorporating these methods into our behavioural model can benefit the accuracy improvement.

\section*{Acknowledgment}
We wish to acknowledge the support from the National Natural Science Foundation of China (Grant No. 72101276).



\section*{References}

\bibliography{main1}

\begin{thebibliography}{28}
\expandafter\ifx\csname natexlab\endcsname\relax\def\natexlab#1{#1}\fi
\providecommand{\url}[1]{\texttt{#1}}
\providecommand{\href}[2]{#2}
\providecommand{\path}[1]{#1}
\providecommand{\DOIprefix}{doi:}
\providecommand{\ArXivprefix}{arXiv:}
\providecommand{\URLprefix}{URL: }
\providecommand{\Pubmedprefix}{pmid:}
\providecommand{\doi}[1]{\href{http://dx.doi.org/#1}{\path{#1}}}
\providecommand{\Pubmed}[1]{\href{pmid:#1}{\path{#1}}}
\providecommand{\bibinfo}[2]{#2}
\ifx\xfnm\relax \def\xfnm[#1]{\unskip,\space#1}\fi
\bibitem[{Alexandre et~al.(2016)Alexandre, Kratarth, Vignesh, Alexandre, Li \&
  Silvio}]{Alexandre2016}
\bibinfo{author}{Alexandre, A.}, \bibinfo{author}{Kratarth, G.},
  \bibinfo{author}{Vignesh, R.}, \bibinfo{author}{Alexandre, R.},
  \bibinfo{author}{Li, F.-F.}, \& \bibinfo{author}{Silvio, S.}
  (\bibinfo{year}{2016}).
\newblock \bibinfo{title}{Social lstm: Human trajectory prediction in crowded
  spaces}.
\newblock In {\it \bibinfo{booktitle}{Proceedings of the IEEE Conference on
  Computer Vision and Pattern Recognition}\/} (pp. \bibinfo{pages}{961--971}).
\bibitem[{Bain \& Bartolo(2019)}]{Bain2019}
\bibinfo{author}{Bain, N.}, \& \bibinfo{author}{Bartolo, D.}
  (\bibinfo{year}{2019}).
\newblock \bibinfo{title}{Dynamic response and hydrodynamics of polarized
  crowds}.
\newblock {\it \bibinfo{journal}{Science}\/},  {\it \bibinfo{volume}{363}\/},
  \bibinfo{pages}{46--49}.
\bibitem[{Ballerini et~al.(2008)Ballerini, Cabibbo, Candelier, Cavagna,
  Cisbani, Giardina, Lecomte, Orlandi, Parisi, Procaccini, Viale \&
  Zdravkovic}]{Ballerini2008}
\bibinfo{author}{Ballerini, M.}, \bibinfo{author}{Cabibbo, N.},
  \bibinfo{author}{Candelier, R.}, \bibinfo{author}{Cavagna, A.},
  \bibinfo{author}{Cisbani, E.}, \bibinfo{author}{Giardina, I.},
  \bibinfo{author}{Lecomte, V.}, \bibinfo{author}{Orlandi, A.},
  \bibinfo{author}{Parisi, G.}, \bibinfo{author}{Procaccini, A.},
  \bibinfo{author}{Viale, M.}, \& \bibinfo{author}{Zdravkovic, V.}
  (\bibinfo{year}{2008}).
\newblock \bibinfo{title}{Interaction ruling animal collective behavior depends
  on topological rather than metric distance: evidence from a field study}.
\newblock {\it \bibinfo{journal}{Proceedings of the National Academy of
  Sciences of the United States of America}\/},  {\it \bibinfo{volume}{105}\/},
  \bibinfo{pages}{1232--7}.
\bibitem[{Blue \& Adler(2001)}]{Blue2001}
\bibinfo{author}{Blue, V.~J.}, \& \bibinfo{author}{Adler, J.~L.}
  (\bibinfo{year}{2001}).
\newblock \bibinfo{title}{Cellular automata microsimulation for modeling
  bi-directional pedestrian walkways}.
\newblock {\it \bibinfo{journal}{Transportation Research Part
  B-Methodological}\/},  {\it \bibinfo{volume}{35}\/},
  \bibinfo{pages}{293--312}.
\bibitem[{Boltes \& Seyfried(2013)}]{Boltes2013}
\bibinfo{author}{Boltes, M.}, \& \bibinfo{author}{Seyfried, A.}
  (\bibinfo{year}{2013}).
\newblock \bibinfo{title}{Collecting pedestrian trajectories}.
\newblock {\it \bibinfo{journal}{Neurocomputing}\/},  {\it
  \bibinfo{volume}{100}\/}, \bibinfo{pages}{127--133}.
\bibitem[{Boltes et~al.(2010)Boltes, Seyfried, Steffen \&
  Schadschneider}]{Boltes2010}
\bibinfo{author}{Boltes, M.}, \bibinfo{author}{Seyfried, A.},
  \bibinfo{author}{Steffen, B.}, \& \bibinfo{author}{Schadschneider, A.}
  (\bibinfo{year}{2010}).
\newblock \bibinfo{title}{Automatic extraction of pedestrian trajectories from
  video recordings}.
\newblock {\it \bibinfo{journal}{Pedestrian and Evacuation Dynamics 2008}\/},
  (pp. \bibinfo{pages}{43--54}).
\bibitem[{Boltes et~al.(2019)Boltes, Zhang, Tordeux, Schadschneider \&
  Seyfried}]{Boltes2019}
\bibinfo{author}{Boltes, M.}, \bibinfo{author}{Zhang, J.},
  \bibinfo{author}{Tordeux, A.}, \bibinfo{author}{Schadschneider, A.}, \&
  \bibinfo{author}{Seyfried, A.} (\bibinfo{year}{2019}).
\newblock \bibinfo{title}{Empirical results of pedestrian and evacuation
  dynamics}.
\newblock {\it \bibinfo{journal}{Complex Dynamics of Traffic Management}\/},
  (pp. \bibinfo{pages}{671--699}).
\bibitem[{Douthwaite et~al.(2019)Douthwaite, Zhao \&
  Mihaylova}]{Douthwaite2019}
\bibinfo{author}{Douthwaite, J.~A.}, \bibinfo{author}{Zhao, S.}, \&
  \bibinfo{author}{Mihaylova, L.~S.} (\bibinfo{year}{2019}).
\newblock \bibinfo{title}{Velocity obstacle approaches for multi-agent
  collision avoidance}.
\newblock {\it \bibinfo{journal}{Unmanned Systems}\/},  {\it
  \bibinfo{volume}{7}\/}, \bibinfo{pages}{55--64}.
\bibitem[{Fiorini \& Shiller(1998)}]{Fiorini1998}
\bibinfo{author}{Fiorini, P.}, \& \bibinfo{author}{Shiller, Z.}
  (\bibinfo{year}{1998}).
\newblock \bibinfo{title}{Motion planning in dynamic environments using
  velocity obstacles}.
\newblock {\it \bibinfo{journal}{The International Journal of Robotics
  Research}\/},  {\it \bibinfo{volume}{17}\/}, \bibinfo{pages}{760--772}.
\bibitem[{Fortune(1987)}]{Fortune1987}
\bibinfo{author}{Fortune, S.} (\bibinfo{year}{1987}).
\newblock \bibinfo{title}{A sweepline algorithm for voronoi diagrams}.
\newblock {\it \bibinfo{journal}{Algorithmica}\/},  {\it
  \bibinfo{volume}{2}\/}, \bibinfo{pages}{153--174}.
\bibitem[{Fu et~al.(2018)Fu, Jia, Chen, Ma, Han \& Luo}]{Fu2018}
\bibinfo{author}{Fu, Z.}, \bibinfo{author}{Jia, Q.}, \bibinfo{author}{Chen,
  J.}, \bibinfo{author}{Ma, J.}, \bibinfo{author}{Han, K.}, \&
  \bibinfo{author}{Luo, L.} (\bibinfo{year}{2018}).
\newblock \bibinfo{title}{A fine discrete field cellular automaton for
  pedestrian dynamics integrating pedestrian heterogeneity, anisotropy, and
  time-dependent characteristics}.
\newblock {\it \bibinfo{journal}{Transportation Research Part C: Emerging
  Technologies}\/},  {\it \bibinfo{volume}{91}\/}, \bibinfo{pages}{37--61}.
\bibitem[{Hart et~al.(1968)Hart, Nilsson \& Raphael}]{Hart1968}
\bibinfo{author}{Hart, P.~E.}, \bibinfo{author}{Nilsson, N.~J.}, \&
  \bibinfo{author}{Raphael, B.} (\bibinfo{year}{1968}).
\newblock \bibinfo{title}{A formal basis for the heuristic determination of
  minimum cost paths}.
\newblock {\it \bibinfo{journal}{IEEE transactions on Systems Science
  Cybernetics}\/},  {\it \bibinfo{volume}{4}\/}, \bibinfo{pages}{100--107}.
\bibitem[{Helbing et~al.(2005)Helbing, Buzna, Johansson \&
  Werner}]{Helbing2005}
\bibinfo{author}{Helbing, D.}, \bibinfo{author}{Buzna, L.},
  \bibinfo{author}{Johansson, A.}, \& \bibinfo{author}{Werner, T.}
  (\bibinfo{year}{2005}).
\newblock \bibinfo{title}{Self-organized pedestrian crowd dynamics:
  Experiments, simulations, and design solutions}.
\newblock {\it \bibinfo{journal}{Transportation Science}\/},  {\it
  \bibinfo{volume}{39}\/}, \bibinfo{pages}{1--24}.
\bibitem[{Helbing et~al.(2000)Helbing, Farkas \& Vicsek}]{Helbing2000}
\bibinfo{author}{Helbing, D.}, \bibinfo{author}{Farkas, I.}, \&
  \bibinfo{author}{Vicsek, T.} (\bibinfo{year}{2000}).
\newblock \bibinfo{title}{Simulating dynamical features of escape panic}.
\newblock {\it \bibinfo{journal}{Nature}\/},  {\it \bibinfo{volume}{407}\/},
  \bibinfo{pages}{487--90}.
\bibitem[{Helbing \& Johansson(2009)}]{Helbing2009}
\bibinfo{author}{Helbing, D.}, \& \bibinfo{author}{Johansson, A.}
  (\bibinfo{year}{2009}).
\newblock \bibinfo{title}{Pedestrian, crowd and evacuation dynamics}.
\newblock In {\it \bibinfo{booktitle}{Encyclopedia of complexity and systems
  science}\/} (pp. \bibinfo{pages}{6476--6495}).
\newblock \bibinfo{publisher}{Springer}.
\bibitem[{Helbing \& Molnar(1995)}]{Helbing1995}
\bibinfo{author}{Helbing, D.}, \& \bibinfo{author}{Molnar, P.}
  (\bibinfo{year}{1995}).
\newblock \bibinfo{title}{Social force model for pedestrian dynamics}.
\newblock {\it \bibinfo{journal}{Physical Review E}\/},  {\it
  \bibinfo{volume}{51}\/}, \bibinfo{pages}{4282--4286}.
\bibitem[{Khambhaita \& Alami(2020)}]{Khambhaita2020}
\bibinfo{author}{Khambhaita, H.}, \& \bibinfo{author}{Alami, R.}
  (\bibinfo{year}{2020}).
\newblock \bibinfo{title}{Viewing robot navigation in human environment as a
  cooperative activity}.
\newblock In \bibinfo{editor}{N.~M. Amato}, \bibinfo{editor}{G.~Hager},
  \bibinfo{editor}{S.~Thomas}, \& \bibinfo{editor}{M.~Torres-Torriti} (Eds.),
  {\it \bibinfo{booktitle}{Robotics Research}\/} (pp.
  \bibinfo{pages}{285--300}).
\newblock \bibinfo{publisher}{Springer International Publishing}.
\bibitem[{Ma et~al.(2019)Ma, Lee, Hu, Shi \& Yuen}]{Ma2019}
\bibinfo{author}{Ma, Y.}, \bibinfo{author}{Lee, E.~W.}, \bibinfo{author}{Hu,
  Z.}, \bibinfo{author}{Shi, M.}, \& \bibinfo{author}{Yuen, R.~K.}
  (\bibinfo{year}{2019}).
\newblock \bibinfo{title}{An intelligence-based approach for prediction of
  microscopic pedestrian walking behavior}.
\newblock {\it \bibinfo{journal}{IEEE Transactions on Intelligent
  Transportation Systems}\/},  (pp. \bibinfo{pages}{1--17}).
\bibitem[{Moussaid et~al.(2011)Moussaid, Helbing \& Theraulaz}]{Moussaid2011}
\bibinfo{author}{Moussaid, M.}, \bibinfo{author}{Helbing, D.}, \&
  \bibinfo{author}{Theraulaz, G.} (\bibinfo{year}{2011}).
\newblock \bibinfo{title}{How simple rules determine pedestrian behavior and
  crowd disasters}.
\newblock {\it \bibinfo{journal}{Proceedings of the National Academy of
  Sciences of the United States of America}\/},  {\it \bibinfo{volume}{108}\/},
  \bibinfo{pages}{6884--8}.
\bibitem[{Older(1968)}]{Older1968}
\bibinfo{author}{Older, S.} (\bibinfo{year}{1968}).
\newblock {\it \bibinfo{title}{Movement of pedestrians on footways in shopping
  streets}\/}.
\newblock \bibinfo{publisher}{Traffic engineering and control}.
\bibitem[{Pandey et~al.(2017)Pandey, Pandey \& Parhi}]{Pandey2017}
\bibinfo{author}{Pandey, A.}, \bibinfo{author}{Pandey, S.}, \&
  \bibinfo{author}{Parhi, D.} (\bibinfo{year}{2017}).
\newblock \bibinfo{title}{Mobile robot navigation and obstacle avoidance
  techniques: A review}.
\newblock {\it \bibinfo{journal}{Int Rob Auto J}\/},  {\it
  \bibinfo{volume}{2}\/}, \bibinfo{pages}{00022}.
\bibitem[{Seyfried et~al.(2005)Seyfried, Steffen, Klingsch \&
  Boltes}]{Seyfried2005}
\bibinfo{author}{Seyfried, A.}, \bibinfo{author}{Steffen, B.},
  \bibinfo{author}{Klingsch, W.}, \& \bibinfo{author}{Boltes, M.}
  (\bibinfo{year}{2005}).
\newblock \bibinfo{title}{The fundamental diagram of pedestrian movement
  revisited}.
\newblock {\it \bibinfo{journal}{Journal of Statistical Mechanics: Theory and
  Experiment}\/},  {\it \bibinfo{volume}{2005}\/}, \bibinfo{pages}{P10002}.
\bibitem[{Van Den~Berg et~al.(2011)Van Den~Berg, Guy, Lin \&
  Manocha}]{VanDenBerg2011}
\bibinfo{author}{Van Den~Berg, J.}, \bibinfo{author}{Guy, S.~J.},
  \bibinfo{author}{Lin, M.}, \& \bibinfo{author}{Manocha, D.}
  (\bibinfo{year}{2011}).
\newblock \bibinfo{title}{Reciprocal n-body collision avoidance}.
\newblock In {\it \bibinfo{booktitle}{Robotics research}\/} (pp.
  \bibinfo{pages}{3--19}).
\newblock \bibinfo{publisher}{Springer}.
\bibitem[{Weidmann(1993)}]{Weidmann1993}
\bibinfo{author}{Weidmann, U.} (\bibinfo{year}{1993}).
\newblock {\it \bibinfo{title}{Transporttechnik der Fussgänger:
  Transporttechnische Eigenschaften des Fussgängerverkehrs
  (Literaturauswertung)}\/}.
\newblock \bibinfo{publisher}{ETH, IVT}.
\bibitem[{Xiao et~al.(2018)Xiao, Chraibi, Qu, Tordeux \& Gao}]{Xiao2018}
\bibinfo{author}{Xiao, Y.}, \bibinfo{author}{Chraibi, M.}, \bibinfo{author}{Qu,
  Y.}, \bibinfo{author}{Tordeux, A.}, \& \bibinfo{author}{Gao, Z.}
  (\bibinfo{year}{2018}).
\newblock \bibinfo{title}{Investigation of voronoi diagram based direction
  choices using uni- and bi-directional trajectory data}.
\newblock {\it \bibinfo{journal}{Physical Review E}\/},  {\it
  \bibinfo{volume}{97}\/}, \bibinfo{pages}{052127}.
\bibitem[{Xiao et~al.(2016)Xiao, Gao, Qu \& Li}]{Xiao2016}
\bibinfo{author}{Xiao, Y.}, \bibinfo{author}{Gao, Z.~Y.}, \bibinfo{author}{Qu,
  Y.~C.}, \& \bibinfo{author}{Li, X.~G.} (\bibinfo{year}{2016}).
\newblock \bibinfo{title}{A pedestrian flow model considering the impact of
  local density: Voronoi diagram based heuristics approach}.
\newblock {\it \bibinfo{journal}{Transportation Research Part C-Emerging
  Technologies}\/},  {\it \bibinfo{volume}{68}\/}, \bibinfo{pages}{566--580}.
\bibitem[{Xiao et~al.(2022)Xiao, Xu, Chraibi, Zhang \& Gou}]{Xiao2022}
\bibinfo{author}{Xiao, Y.}, \bibinfo{author}{Xu, J.}, \bibinfo{author}{Chraibi,
  M.}, \bibinfo{author}{Zhang, J.}, \& \bibinfo{author}{Gou, C.}
  (\bibinfo{year}{2022}).
\newblock \bibinfo{title}{A generalized trajectories-based evaluation approach
  for pedestrian evacuation models}.
\newblock {\it \bibinfo{journal}{Safety Science}\/},  {\it
  \bibinfo{volume}{147}\/}, \bibinfo{pages}{105574}.
\bibitem[{Zhang et~al.(2019)Zhang, Ouyang, Zhang, Xue \& Zheng}]{Zhang2019}
\bibinfo{author}{Zhang, P.}, \bibinfo{author}{Ouyang, W.},
  \bibinfo{author}{Zhang, P.}, \bibinfo{author}{Xue, J.}, \&
  \bibinfo{author}{Zheng, N.} (\bibinfo{year}{2019}).
\newblock \bibinfo{title}{Sr-lstm: State refinement for lstm towards pedestrian
  trajectory prediction}.
\newblock In {\it \bibinfo{booktitle}{Proceedings of the IEEE Conference on
  Computer Vision and Pattern Recognition}\/} (pp.
  \bibinfo{pages}{12085--12094}).

\end{thebibliography}
\bibliographystyle{style}\biboptions{authoryear}

\newpage
\newpage

\begin{appendix}
\setcounter{figure}{0}

\section{Detour coefficient} \label{section appendix detour coefficient}

The detour coefficient $\sigma_i$ is meant to indicate the difficulty level of the future journey to the destination, and here it is indicated according to the potential conflicts using a geometric approach. The approach can be described with two steps as follows.

\begin{figure}[!ht]
\centering{\includegraphics[width=0.3\textwidth]{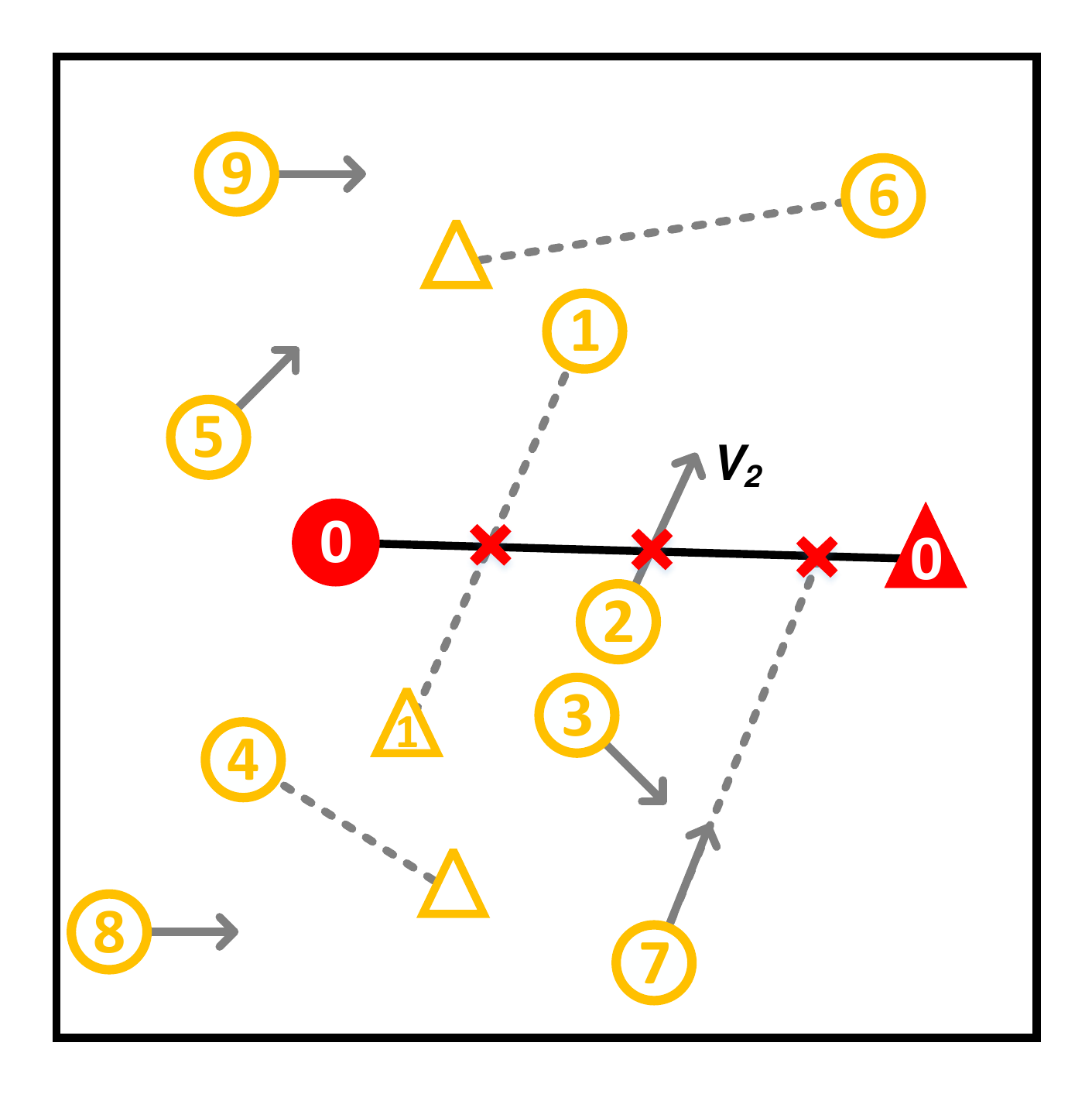}}
\caption{Sketch map of the estimated detour degree.} \label{figdetourdegree}
\end{figure}

\textbf{Step 1: The identification of potential conflicting point.} For the object pedestrian $P_0$ (see Fig. \ref{figdetourdegree}), the neighboring pedestrians are divided into two types depending on that whether the destination is clear or not. For the first type of pedestrian whose destination is clear, the potential conflicting point is identified by judging the intersection between two segments from the pedestrian to the destination (e.g., segment $\overline{P_0D_0}$ and segment $\overline{P_1D_1}$). For the second type whose destination is unclear for the object pedestrian, the potential conflicting point is the intersection between the segment of object pedestrian and the velocity ray of the related pedestrian (e.g., segment $\overline{P_0D_0}$ and ray $\bm{v}_2$). Note that in many cases no potential conflicting point is identified when there is no intersection point.

\textbf{Step 2: The calculation of potential conflict.} For each conflicting point, the conflicting time gap is calculated by $\Delta{t_{ij}} =  \left|d_i^j/v_i - d_j^i/v_j \right|$, where $d_i^j$ and $d_j^i$ respectively denotes the distance of pedestrian $P_i$ and $P_j$ to the conflicting point. Accordingly, the detour coefficient $\sigma_i$ is calculated as,  
\begin{equation}\label{eq detour degree}
\sigma_i = \sum_j{e^{-\Delta{t_{ij}}}}/D_i,
\end{equation}
where $D_i$ indicates the distance between pedestrian $P_i$ to the corresponding destination.

\end{appendix}

\end{document}